\documentclass[aps,pra,groupedaddress,superscriptaddress,amsmath,amssymb,showpacs,twocolumn]{revtex4-1}
\usepackage{graphicx}
\usepackage{bm,bbm,braket}
\usepackage{color}
\usepackage{times,txfonts}
\usepackage{hyperref}

\usepackage[caption=false]{subfig}

\def\Tr{\mathrm{Tr}}
\DeclareMathOperator{\SL}{SL}
\begin{document}

\title{Strong monogamy inequalities for four qubits}

\author{Bartosz Regula}
\affiliation{$\mbox{School of Mathematical Sciences, The University of Nottingham, University Park,
Nottingham NG7 2RD, United Kingdom}$}

\author{Andreas Osterloh}
\affiliation{Institut f\"{u}r Theoretische Physik, Universit\"{a}t Duisburg-Essen, D-47048 Duisburg, Germany}

\author{Gerardo Adesso}
\affiliation{$\mbox{School of Mathematical Sciences, The University of Nottingham, University Park,
Nottingham NG7 2RD, United Kingdom}$}

\date{\today}

\begin{abstract}
We investigate possible generalizations of the Coffman-Kundu-Wootters monogamy inequality to four qubits, accounting for multipartite entanglement in addition to the bipartite terms. We show that the most natural extension of the inequality does not hold in general, and we describe the violations of this inequality in detail. We investigate alternative ways to extend the monogamy inequality to express a constraint on entanglement sharing valid for all four-qubit states, and perform an extensive numerical analysis of randomly generated four-qubit states to explore the properties of such extensions.
\end{abstract}


\maketitle

\section{Introduction}

The conceptual and computational difficulties of quantifying the entanglement of larger systems make the complete description of multipartite entanglement one of the biggest challenges of quantum information theory \cite{Horodecki2009,eltschka_2014-1}. Gaining a better insight into the properties of entangled states of many parties would lead to major progress in our understanding of the concept of entanglement, and would have significant implications on many quantum information protocols and other physical processes which rely on it \cite{nielsen_2011,amico_2008}.

One of the fundamental traits of multipartite quantum entanglement is its {\it monogamy} --- an inherent constraint on how it can be shared among multiple parties. The concept was first formalized for a system of three qubits in the seminal work of Coffman, Kundu, and Wootters (CKW) \cite{coffman_2000}, showing that the entanglement of a qubit with another pair of qubits (quantified with the \emph{tangle} $\tau$) places a bound on the total amount of pairwise entanglement between the considered qubit and each qubit from the pair. This relation, referred to as the CKW monogamy inequality, can be expressed for a pure state  $\ket{\phi}$ of three qubits as
\begin{equation}\label{eq:ckw}
\tau_{1|23}^{(1)} \ (\ket\phi) \geq  \tau_{1|2}^{(2)} \ (\ket\phi)+ \tau_{1|3}^{(2)}\ (\ket\phi)
\end{equation}
where $\tau_{i|jk}^{(1)}$ represents the \emph{one-tangle}, also known as the \emph{linear entropy}, quantifying the bipartite entanglement between a chosen qubit $i$ and the rest of the system in the state $\ket\phi$:
\begin{equation}
    \tau_{i|jk}^{(1)} \ (\ket\phi) = 4 \det \left( \Tr_{jk} \ket\phi \bra\phi \right)
\end{equation}
and  $\tau_{i|j}^{(2)}$ denotes the \emph{two-tangle}, which measures the pairwise entanglement for the reduced subsystem of qubits $i$ and $j$:
\begin{equation}
    \tau_{i|j}^{(2)} \ (\ket\phi) = \mathcal{C}^2 \left ( \Tr_{k} \ket\phi \bra\phi \right)\; .
\end{equation}
Here, $\mathcal{C}$ is the concurrence \cite{hill_1997,wootters_1998}.

Remarkably, the residual entanglement in this relation is invariant under stochastic local operations and classical communication (SLOCC), as well as under permutations of the parties $ijk$, and in fact quantifies the genuine three-partite entanglement in the system \cite{dur_2000}. It is is now known as the \emph{three-tangle}:
\begin{equation}\label{eq:threetangle}
\tau_{1|2|3}^{(3)} \ (\ket\phi) = \tau_{1|23}^{(1)} \ (\ket\phi) -  \tau_{1|2}^{(2)} \ (\ket\phi) - \tau_{1|3}^{(2)} \ (\ket\phi)
\end{equation}
and it admits a straightforward closed-form expression as a polynomial of homogeneous degree 4 in the coefficients of the state $\ket\phi$ \cite{coffman_2000, verstraete_2003, osterloh_2005}.

Even though it is tailored for particular choices of the entanglement measure, the CKW monogamy inequality is often regarded as a quantitative constraint capturing a characteristic property of entanglement itself, which distinguishes it from other, weaker forms of nonclassical correlations \cite{streltsov_2012,lancien_2016}. The relation was later extended by Osborne and Verstraete \cite{osborne_2006} to describe the distribution of pairwise entanglement in a system of $n$ qubits, following the conjecture by CKW \cite{coffman_2000}, and takes the form
\begin{equation}\label{eq:osborne}
 \tau_{1 |\, 2 \cdots\, n}^{(1)} \ (\ket\phi) \geq \sum_{j=2}^{n} \tau_{1 | j}^{(2)} \ (\ket\phi).
\end{equation}
It is, however, obvious that this relation does not fully describe the entanglement of an $n$-partite state, since on the right-hand side it only contains the entanglement of pairs of particles and disregards genuine multipartite correlations. Indeed, many attempts have been made to construct a generalized monogamy inequality for $n$ qubits \cite{eltschka_2009,gour_2010-1,cornelio_2013,regula_2014,eltschka_2015}. While in particular \cite{eltschka_2015} established a general symmetric monogamy equality involving various forms of bipartite and multipartite entanglement between different partitions of qubits, no extension in the spirit of inequality \eqref{eq:osborne} was successfully reported to provide a constraint on entanglement sharing which could satisfactorily take into account all the forms of multipartite entanglement involving one focus qubit and subgroups of $2 \leq k \leq n-1$ remaining qubits. In fact, it was shown that such an extension using a generalized form of the tangle based on antilinear operators \cite{osterloh_2005} is impossible \cite{eltschka_2009}.

In this paper, we first investigate an arguably natural generalization of the CKW inequality \eqref{eq:ckw} to four qubits, and show that it cannot hold in the general case by providing analytical counterexamples. We then present other possible methods of extending the inequality, and consider their implications on the validity of a possible general strong monogamy relation by performing an extensive numerical investigation of arbitrary four-qubit states and analyzing some example states in more detail.

\section{Strong monogamy inequalities}

From eqs. \eqref{eq:ckw} and \eqref{eq:osborne}, one might expect a natural extension of the CKW inequality to hold, where the terms corresponding to different kinds of entanglement of up to $(n-1)$ subsystems are considered. However, there are two major questions which make such a generalization non-trivial; namely, how exactly to define and quantify the different types of multipartite entanglement, and whether the entanglement of more parties has to be scaled in the same way as the bipartite entanglement, or if we need some other appropriate scaling. We aim to provide and investigate several alternative answers to these questions.

Let us first consider the simplest generalization of the monogamy inequality for arbitrary pure states  $\ket\phi$ of four qubits, which could take the form (referred to as ``strong monogamy'' \cite{adesso_2007,regula_2014}):
\begin{equation}\label{eq:natural-sm}
    \tau_{1|234}^{(1)} \geq \tau_{1|2}^{(2)} + \tau_{1|3}^{(2)} + \tau_{1|4}^{(2)} + \tau_{1|2|3}^{(3)} + \tau_{1|2|4}^{(3)} + \tau_{1|3|4}^{(3)},
\end{equation}
with $\ket\phi$ omitted for simplicity. Here, the square root of the three-tangle $\tau^{(3)}$ can be extended to mixed states by taking the convex roof \cite{uhlmann_1998}:
\begin{equation}\label{eq:natural-sm2}
    \tau_{i|j|k}^{(3)}(\rho) = \left[ \min_{\{p_n, \ket{\psi_n}\}} \sum_n p_n \sqrt{\tau_{i|j|k}^{(3)}(\ket{\psi_n})} \right]^2
\end{equation}
where the minimum is taken over all convex decompositions of $\rho$ into ensembles of pure states $\{\ket{\psi_n}\}$ with probabilities $\{p_n\}$, $\rho = \sum_n p_n \ket{\psi_n} \bra{\psi_n}$. It is worth noting that although the pure-state three-tangle admits a closed-form expression, the convex roof of the three-tangle is not known in general, and it has been found only for special cases of mixed states \cite{lohmayer_2006,eltschka_2008,jung_2009,siewert_2012,viehmann_2012,eltschka_2012-1,regula_2016}. We also note that the convex roof extension for the three-tangle is not defined unambiguously; what we have written is, in fact, the squared convex roof of the square root of the three-tangle, while other works consider the convex roof of the three-tangle directly. However, the use of the square root of the three-tangle in Eq.~\eqref{eq:natural-sm2} is necessary to obtain a convex roof-extended measure which has homogeneous degree $2$, and therefore scales linearly with the density matrix, generalizing the properties of the pure-state three-tangle \cite{viehmann_2012} and establishing a privileged degree $d=2$ for polynomial measures \cite{eltschka_2014-1}. Every such measure is guaranteed to be an entanglement monotone \cite{verstraete_2003}.

Let us now denote by $\tau_{1|2|3|4}^{(4)}$ the residual term of the inequality, that is,
\begin{equation}\label{eq:residual}
    \tau_{1|2|3|4}^{(4)} = \tau_{1|234}^{(1)} - \tau_{1|2}^{(2)} - \tau_{1|3}^{(2)} - \tau_{1|4}^{(2)} - \tau_{1|2|3}^{(3)} - \tau_{1|2|4}^{(3)} - \tau_{1|3|4}^{(3)},
\end{equation}
which makes Eq.~\eqref{eq:natural-sm} equivalent to $\tau_{1|2|3|4}^{(4)} \geq 0$. Notice that, in general, the residual $\tau_{1|2|3|4}^{(4)}$ is not invariant under permutations of the qubits any more, and has to be considered with respect to a chosen partition which assigns a focus role to qubit $1$. Interestingly, the generalized form of this inequality has been shown to hold in a range of states of $n$ qubits, including mixtures of W states and generalized GHZ states \cite{regula_2014} as well as generalized W states, which saturate the inequality \cite{coffman_2000,eltschka_2008,kim_2014}. It has also been investigated using different entanglement measures, including the square of convex-roof extended negativity \cite{choi_2015} and the standard negativity \cite{karmakar_2016}. Additionally, a constraint equivalent to Eq.~\eqref{eq:natural-sm} has been shown to hold for all permutationally-invariant continuous-variable Gaussian states \cite{adesso_2007}, suggesting that it might be possible to establish a similar relation for qubits.

However, and rather surprisingly, this natural extension turns out not to hold in general, even for four qubits. The violations of this strong monogamy inequality were first conjectured in \cite{regula_2014} based on numerical evidence. Here, we will show more precisely when they occur, using analytical results which allow us to calculate the mixed-state three-tangle in particular settings \cite{regula_2016}. We then explore what alternative extensions of the generalized monogamy inequality are possible to account for these violations.

\subsection{Violations of the inequality}

We begin by noting that while four-qubit states can be divided into infinitely many SLOCC-inequivalent classes \cite{dur_2000,verstraete_2002}, there is, in fact, a convenient way to classify them into nine classes representing nine different ways in which they can be entangled \cite{verstraete_2002, chterental_2007}, where the classes can be subdivided further into a three-dimensional characteristic entanglement vector \cite{osterloh_2005}. The nine classes are defined by SLOCC operations acting on the generating states $\ket{G^i_{a,b,c,d}}$ with $i \in \{1,\ldots,9\}$, where each generator is dependent on at most four continuous complex parameters $a$,$b$,$c$,$d$, and the union of the nine classes covers the whole Hilbert space of pure four-qubit states (up to permutations of the qubits). Therefore, any state $\ket{\phi}$ can be obtained as
\begin{equation}\ket\phi = \big( A_1 \otimes A_2 \otimes A_3 \otimes A_4 \big) \ket{G^i_{a,b,c,d}},\end{equation}
for some choice of the complex parameters $a,b,c,d$ as well as the invertible linear operators $A_k \in \SL(2, \mathbb{C})$ representing SLOCC transformations \cite{dur_2000}, with the index $i$ uniquely determining which SLOCC class the state $\ket\phi$ belongs to.

Here we note a misprint in the classification of \cite{verstraete_2002}, where the generating state $\ket{G^4_{a,b}}$ ($L_{ab_3}$ in the original notation) was reported incorrectly, as first mentioned in \cite{chterental_2007}. The correct expression is actually \cite{spee_2015}
\begin{equation}\begin{aligned}\ket{G^4_{a,b}} &= a\big(\ket{0000}+\ket{1111}\big) + \frac{a+b}{2}\big(\ket{0101}+\ket{1010}\big)\\ & + \frac{a-b}{2}\big(\ket{0110}+\ket{1001}\big)\\
    &+ \frac{i}{\sqrt{2}}\big(-\ket{0001} - \ket{0010} + \ket{0111} + \ket{1011}\big)\end{aligned}\end{equation}
	with the two minus signs in the last line differing from \cite{verstraete_2002}. The original, incorrect $\ket{G^4_{a,b}}$ of \cite{verstraete_2002} is instead equivalent to a degenerate case of SLOCC class 2 \cite{chterental_2007}, which coincidentally happens to be a class leading to violations of the strong monogamy inequality (\ref{eq:natural-sm}), as shown in detail below.
	
To investigate the violations of strong monogamy as defined in eqs. \eqref{eq:natural-sm} and \eqref{eq:natural-sm2}, we first performed extensive numerical tests in search of states whose residual $\tau_{1|2|3|4}^{(4)}$ is negative. Noting that the three-qubit reduced subsystems of each pure four-qubit state are mixed states of rank 2, we remark that (at present) it is not possible in general to have an expression for the convex roof of the three-tangle $\tau_{i|j|k}^{(3)}(\rho)$, and so we employ the so-called zero-$E$ approximation \cite{rodriques_2014} to provide a tight upper bound on the mixed-state three-tangle. This means that our numerical tests in fact bound the residual $\tau_{1|2|3|4}^{(4)}$ from below, and while violations have to be confirmed in more detail (resorting to independent methods), all positive results confirm the inequality. It is not possible to strictly falsify it unless we have an exact result for the convex roof.

A test of $10^6$ randomly generated pure states in each SLOCC class, with all parameters $a$,$b$,$c$,$d$ and invertible linear operators $A_k$ randomized, has returned no violations of the inequality. This provides strong evidence that the natural extension of monogamy given by Eq.~(\ref{eq:natural-sm}) holds for a {\em generic} choice of parameters defining the SLOCC classes. However, the situation is more subtle for non-generic states. Specifically, in class 2, generated by
\begin{equation}\begin{aligned}
	\ket{G^2_{a,b,c}} &= \frac{a+b}{2}\big(\ket{0000}+\ket{1111}\big)+\frac{a-b}{2}\big(\ket{0011}+\ket{1100}\big)\\
	& + c \big(\ket{0101}+\ket{1010}\big) + \ket{0110},\end{aligned}\end{equation}
	we can find degenerate subclasses with $a=c$ or $b=c$ (referred to as the SLOCC class 4 in \cite{verstraete_2002,regula_2014}) which do, in fact, lead to violations of the strong monogamy inequality. A further numerical test of random invertible linear operators applied to $5\times 10^6$ states in each of these degenerate subclasses revealed $24,626$ violations in the former and $24,654$ in the latter, with the most violating state having a residual value of $\tau_{1|2|3|4}^{(4)} \approx -0.097$, which is far beyond any possible error due to numerical precision. As anticipated, violations cannot be conclusively confirmed until one establishes that the adopted upper bound for the mixed-state three-tangles is in fact tight. Fortunately, this is possible in the considered case.

To see this, we note the peculiarity of the two degenerate subclasses of SLOCC class 2. They are a special case where all four three-qubit reduced subsystems of the four-qubit state have only one state with vanishing three-tangle in their ranges. This means that they obey the so-called {\it one-root} property, and the convex roof of the square root of the three-tangle can be obtained \emph{exactly} and has the same value for every convex decomposition \cite{regula_2016}. Therefore, the values of the residual $\tau_{1|2|3|4}^{(4)}$ calculated numerically are in fact the exact values, and any violation of the inequality found in these subclasses is a true violation.

\begin{figure}
    \centering
    \includegraphics[width=8cm]{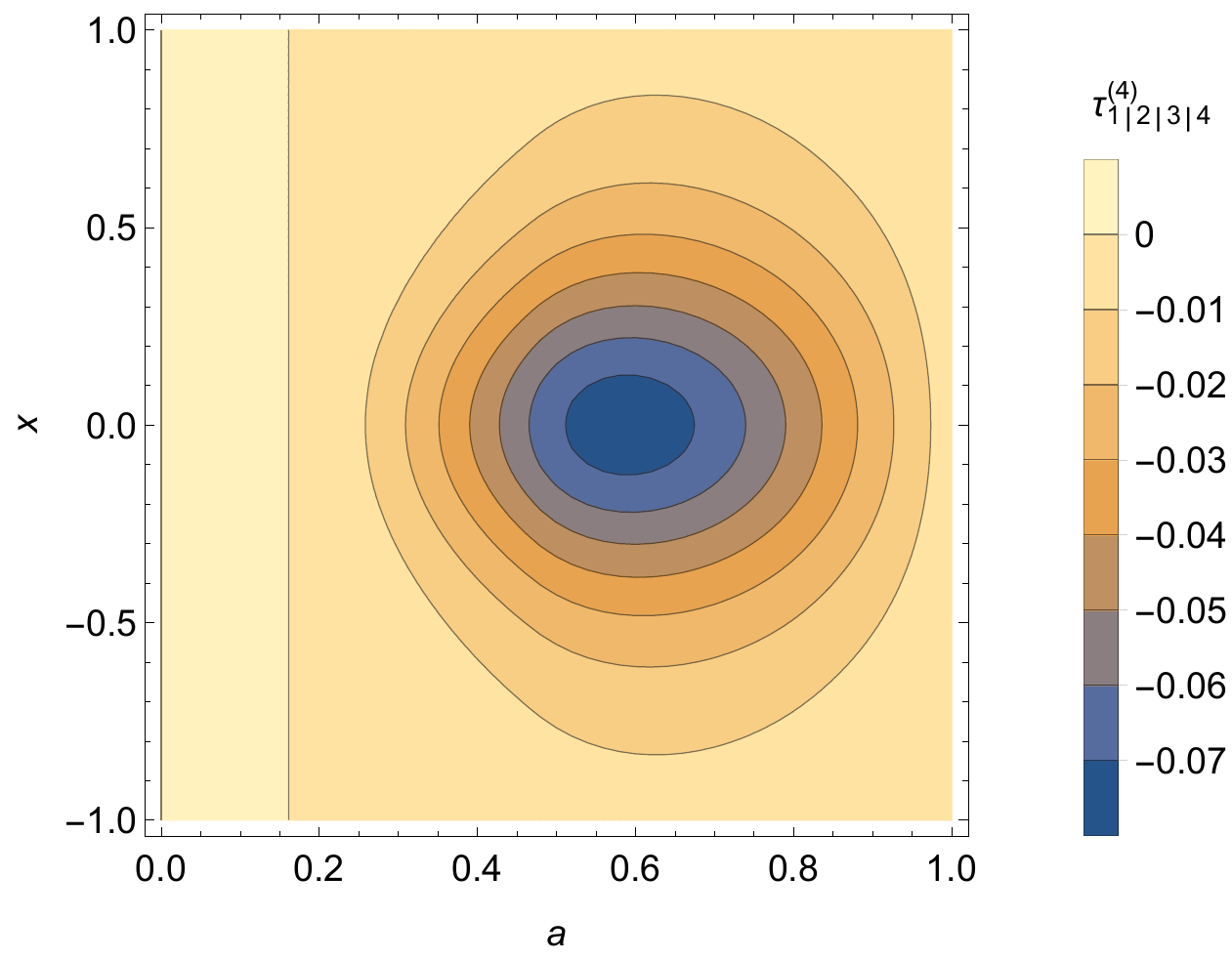}
    \caption{The exact residual $\tau_{1|2|3|4}^{(4)}$ as defined in Eq.~(\ref{eq:residual}) of the state $\rho_{a,x}^1$ (Eq.~\eqref{eq:rho}) with respect to the parameters $a$ and $x$.}
    \label{fig:3dplot}
\end{figure}

In order to present more explicit results, we would like to characterize a family of violations analytically. An example family of states with particularly large violations of the inequality can be constructed by beginning with the state $\ket{G^2_{a,b,c}}$ with $b=c$ and additionally imposing $b=c=ia$ for a parameter $a \geq 0$. We can then choose to parametrize an invertible linear operator $A_x$ of determinant $1$ in a form which leads to an analytically tractable violating state:
\begin{equation}
    A_x = \left(%
\begin{array}{cc}
 (1+i) \left(x+\frac{1}{3}\right) & \left(\frac{1}{3}-i\right)+x \\
 i \left(x-\frac{2}{3}\right) & \frac{1+i}{6}\left(3x-2-3i\right) \\
\end{array}%
\right)
\end{equation}
with a real parameter $x$. The simplest case to consider is to apply the transformation to the first qubit only, giving the state (up to normalization)
\begin{equation}\label{eq:rho}
    \rho_{a,x}^{1} = \Big(A_x \otimes \mathbbm{1} \otimes \mathbbm{1} \otimes \mathbbm{1}\Big) \; \ket{G^2_{a,ia,ia}}\bra{G^2_{a,ia,ia}} \; \Big(A_x \otimes \mathbbm{1} \otimes \mathbbm{1} \otimes \mathbbm{1}\Big)^{\dagger}
\end{equation}
with an explicit dependence on the parameters $a$ and $x$. Since $\rho_{a,x}^1$ belongs to the degenerate subclass of SLOCC class 2 for all $a$ and $x$, we know that we can reach the convex roof of the three-tangles $\tau_{i|j|k}^{(3)}(\rho_{a,x}^1)$ of all reduced subsystems and therefore quantify the residual $\tau_{1|2|3|4}^{(4)}$ of the state $\rho_{a,x}^1$ exactly \cite{regula_2016}. While its explicit expression is lengthy and omitted here, we plot it in Figure \ref{fig:3dplot} for different values of the parameters.

\begin{figure*}
    \centering
    \subfloat[]{\includegraphics[width=5.9cm]{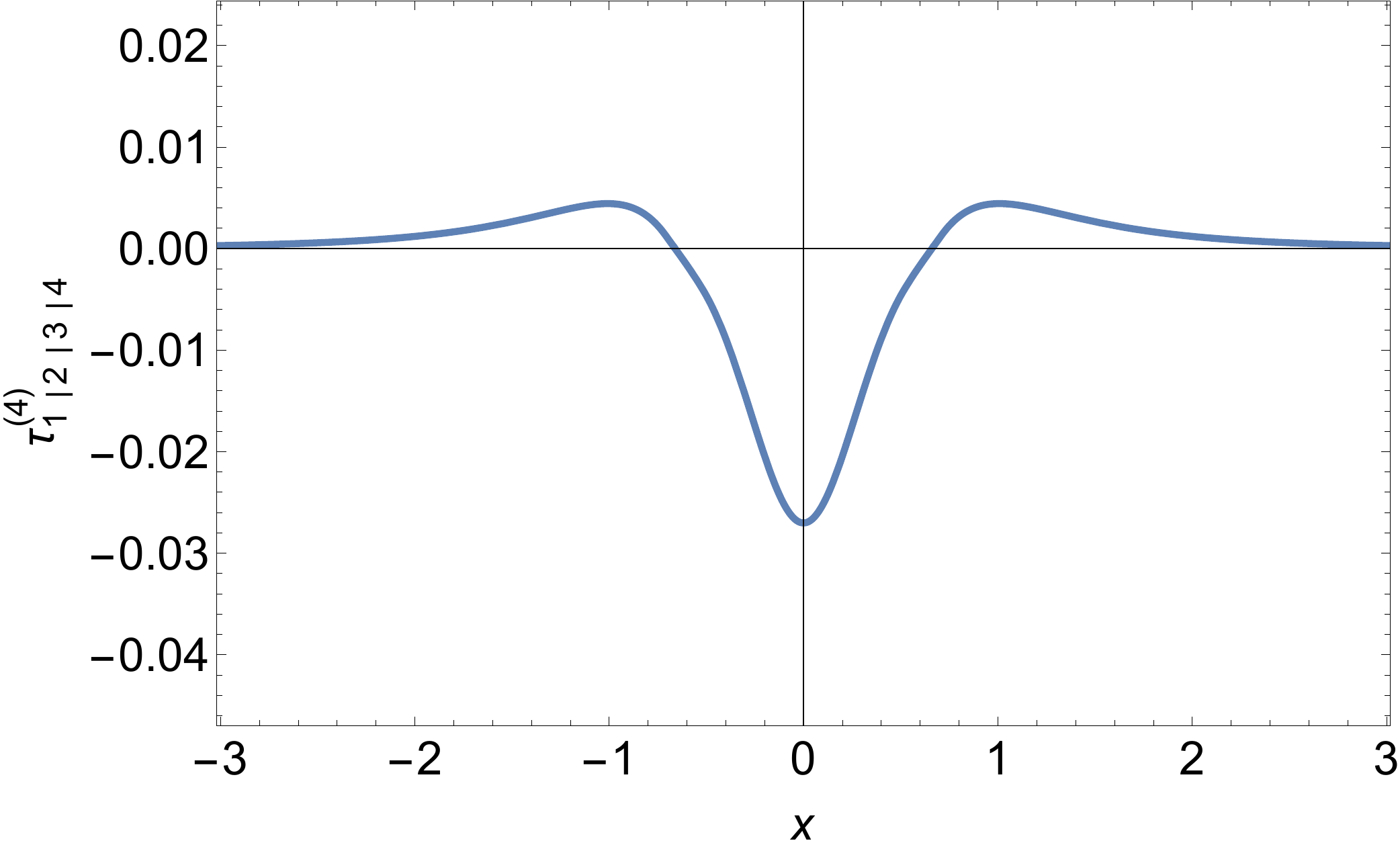}}
    \subfloat[]{\includegraphics[width=5.8cm]{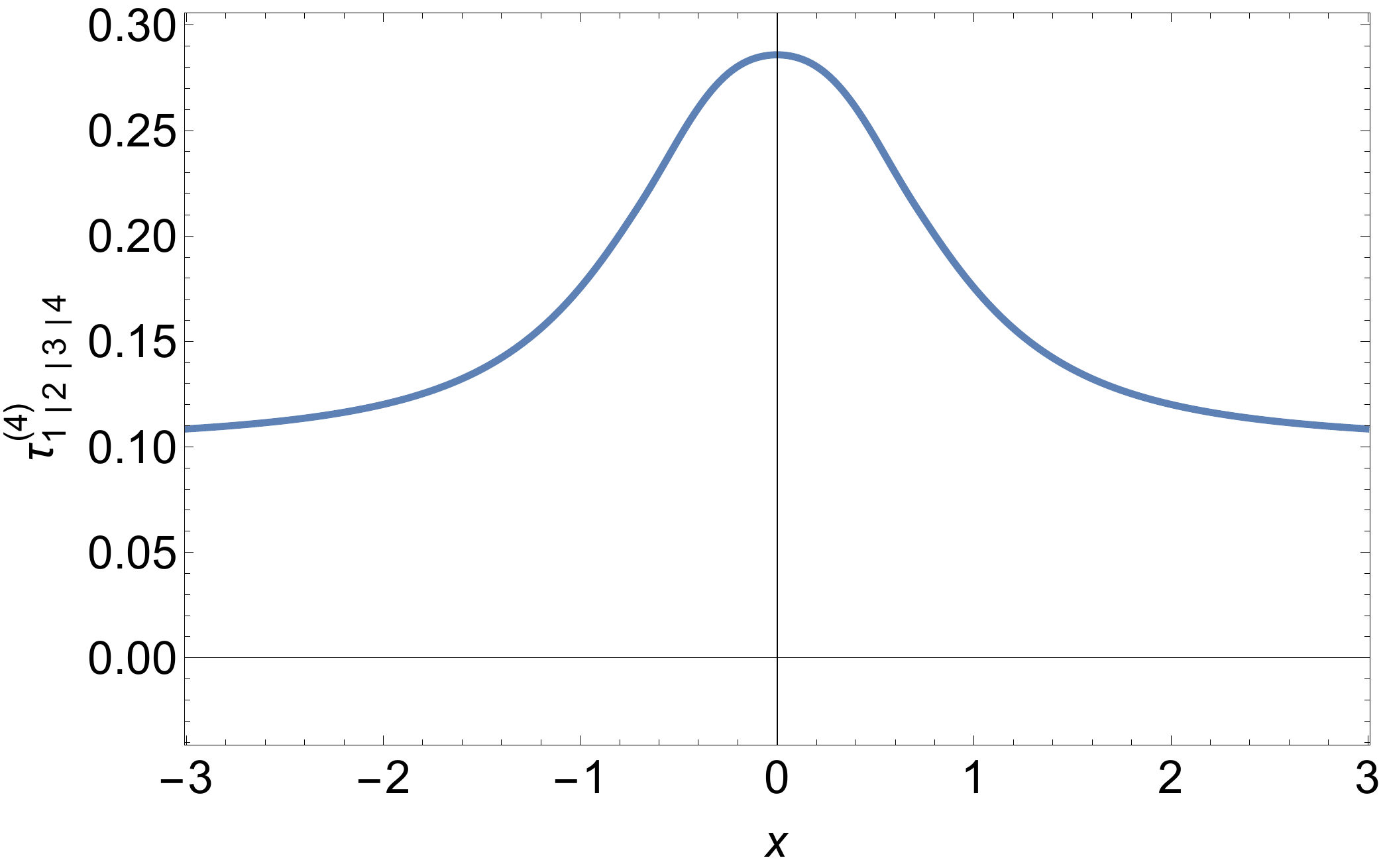}}
    \subfloat[]{\includegraphics[width=5.8cm]{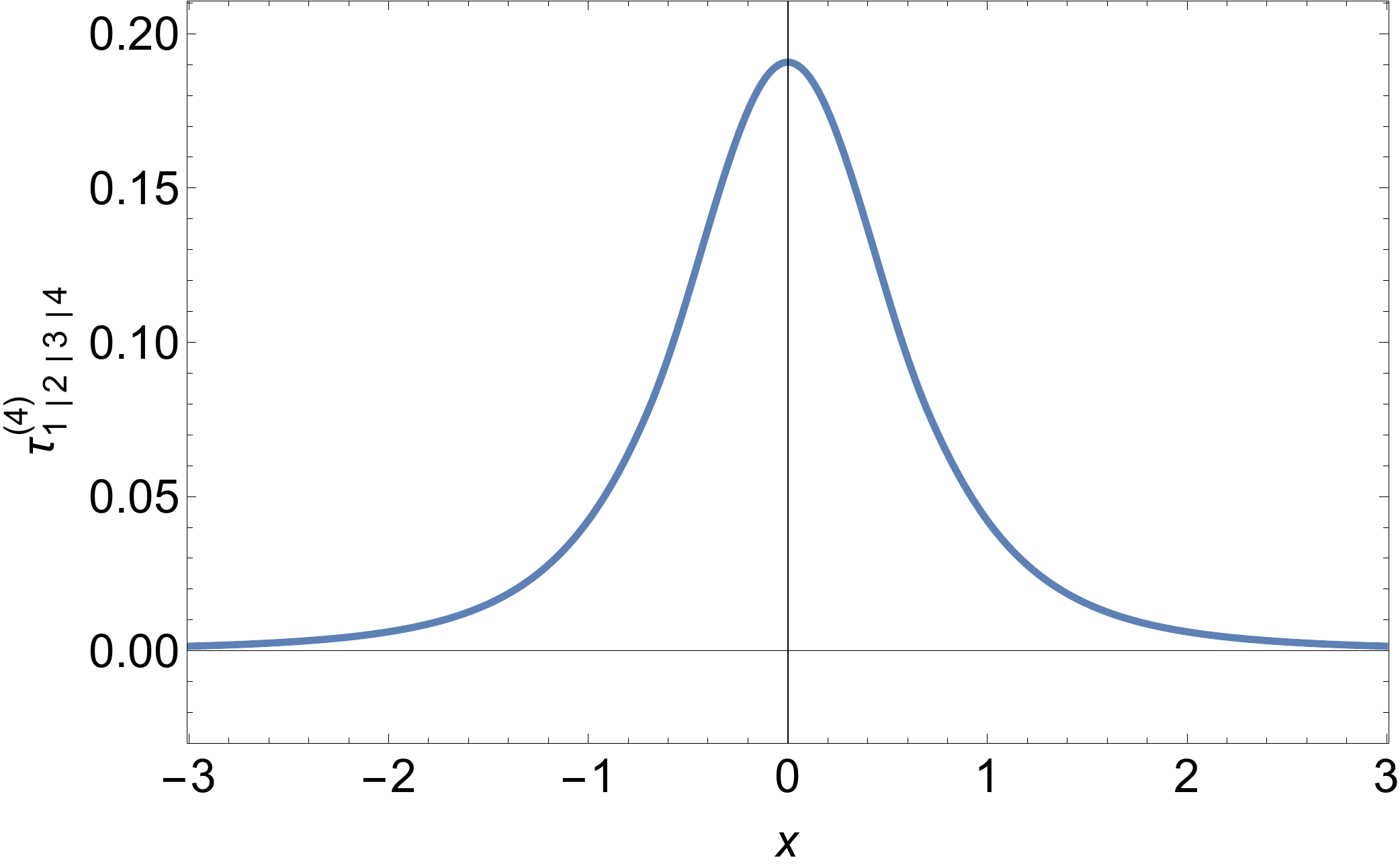}}
    \caption{(a) The exact residual $\tau_{1|2|3|4}^{(4)}$ defined in Eq.~(\ref{eq:residual}) of the state $\rho_{a_0,x}^2$ (see Eq.~\eqref{eq:rho2}).\\*
    (b) The modified residual defined in Eq.~\eqref{eq:mu} of the state $\rho_{a_0,x}^2$, for $\mu = 3$.\\*
    (c) The lower bound for the modified residual defined in Eq.~\eqref{eq:q} of the state $\rho_{a_0,x}^2$, for $q = 4$.}
    \label{fig:res}
\end{figure*}

As can be seen from the figure, the strong monogamy inequality is violated for a large subset of values of $a$ and $x$. One can easily notice that the SLOCC transformation with $x=0$ leads to the biggest violations. To obtain the minimum with regards to $a$, we analyze the marginal one-, two-, and three-tangles of the subsystems, and find that the value of $a =a_0 = \frac{5}{6\sqrt{2}}$ leads in fact to the minimum value of the residual $\tau_{1|2|3|4}^{(4)} \approx -0.076$ within the considered family of states.

We have thus explicitly demonstrated a significant violation of the strong monogamy inequality. This shows that the natural extension of the CKW monogamy inequality in Eqs.~\eqref{eq:natural-sm} and \eqref{eq:natural-sm2} does not hold in general. We then need to consider different forms of the inequality in order to obtain a result which might allow for a relation applicable to to all four-qubit states.

To be able to compare how the different approaches affect the values of the residual, we will use an example state from the same generating family as before, where we fix $a=a_0 = \frac{5}{6\sqrt{2}}$ and apply now the SLOCC transformation $A_x$ to the second qubit only, giving
\begin{equation}\label{eq:rho2}
    \rho_{a_0,x}^{2} = \Big(\mathbbm{1} \otimes A_x \otimes \mathbbm{1} \otimes \mathbbm{1}\Big) \; \ket{G^2_{a_0,i a_0,i a_0}}\bra{G^2_{a_0,i a_0,i a_0}} \; \Big(\mathbbm{1} \otimes A_x \otimes \mathbbm{1} \otimes \mathbbm{1}\Big)^{\dagger}.
\end{equation}
The exact residual of this state for different values of $x$, as defined in Eq.~(\ref{eq:residual}), is plotted in Fig.~\ref{fig:res}(a).

\subsection{Modifying the inequality: approach 1}

The first alternative approach we consider is a rescaling of the three-qubit terms of the inequality with a parameter $\mu \geq 1$:
\begin{equation}
\label{eq:mu}
    \tau_{i|j|k}^{(3)}(\rho) = \left[ \min_{\{p_n, \ket{\psi_n}\}} \sum_n p_n \sqrt{\tau_{i|j|k}^{(3)}(\ket{\psi_n})} \right]^{\ \mu}.
\end{equation}
The meaning of this rescaling is to assigning a lower weight to the three-partite terms compared to the bipartite terms in the monogamy decomposition.

\begin{figure*}
    \centering
    \subfloat[]{\includegraphics[width=5.9cm]{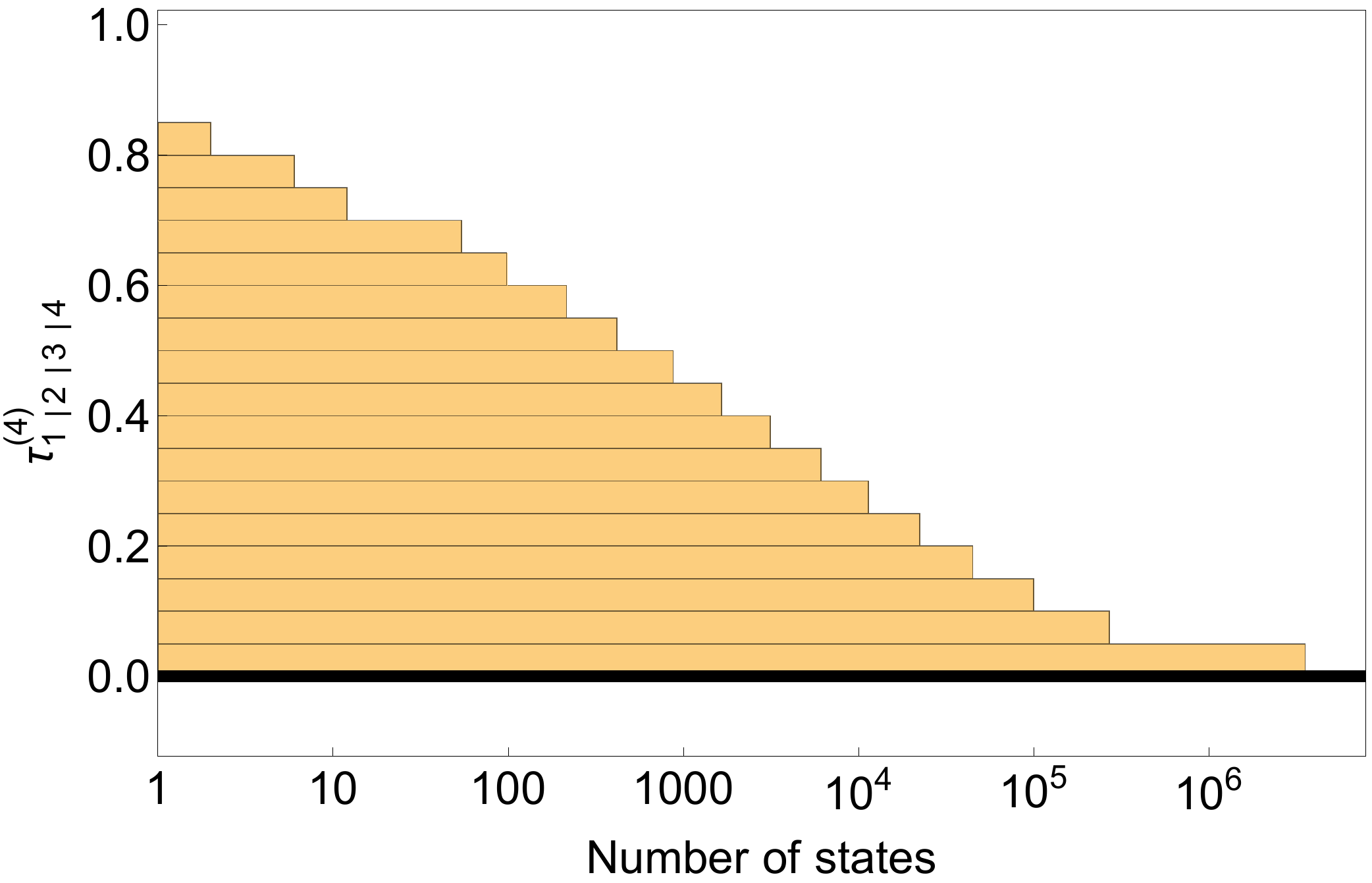}}
    \subfloat[]{\includegraphics[width=5.8cm]{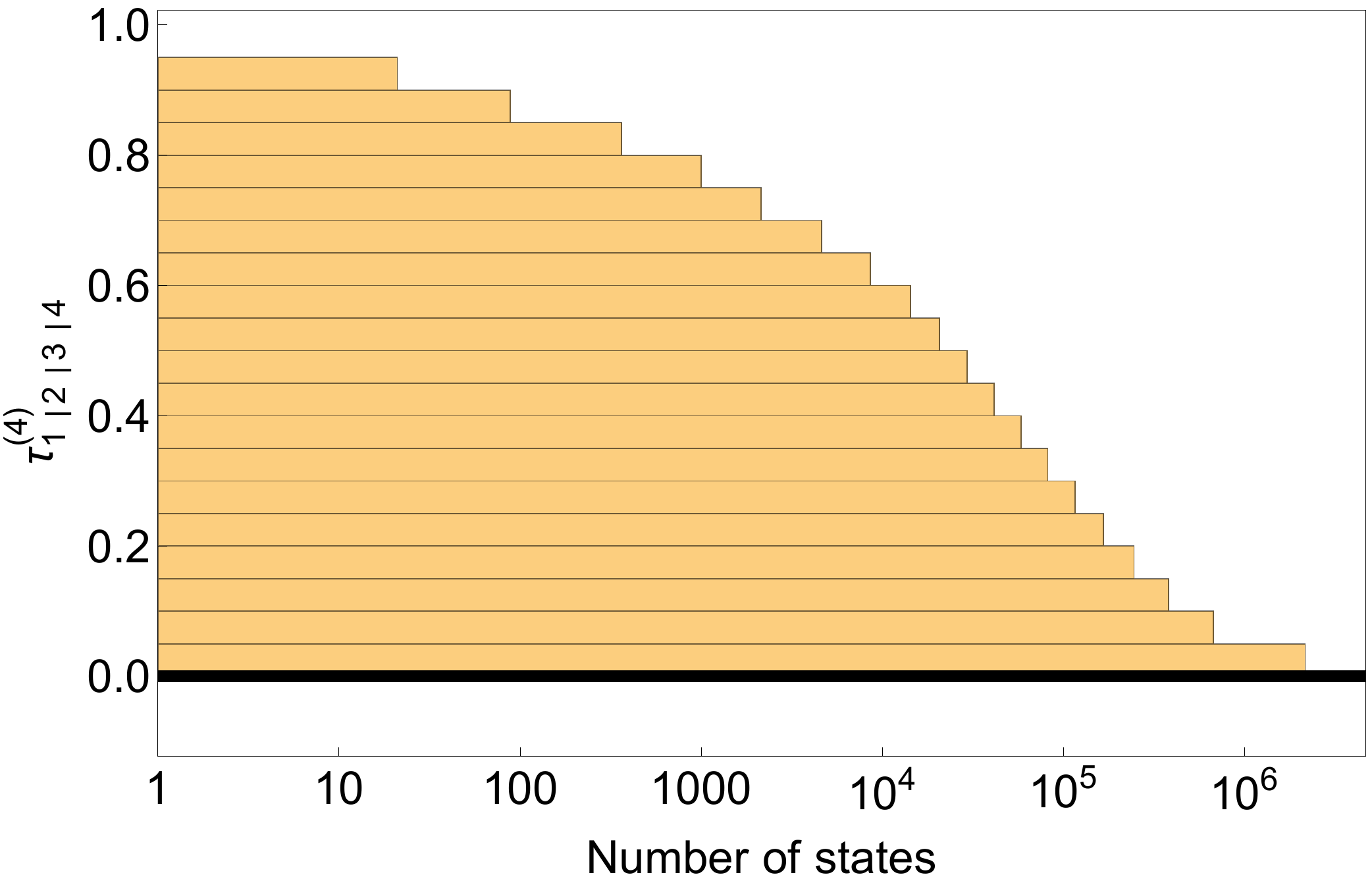}}
    \subfloat[]{\includegraphics[width=5.8cm]{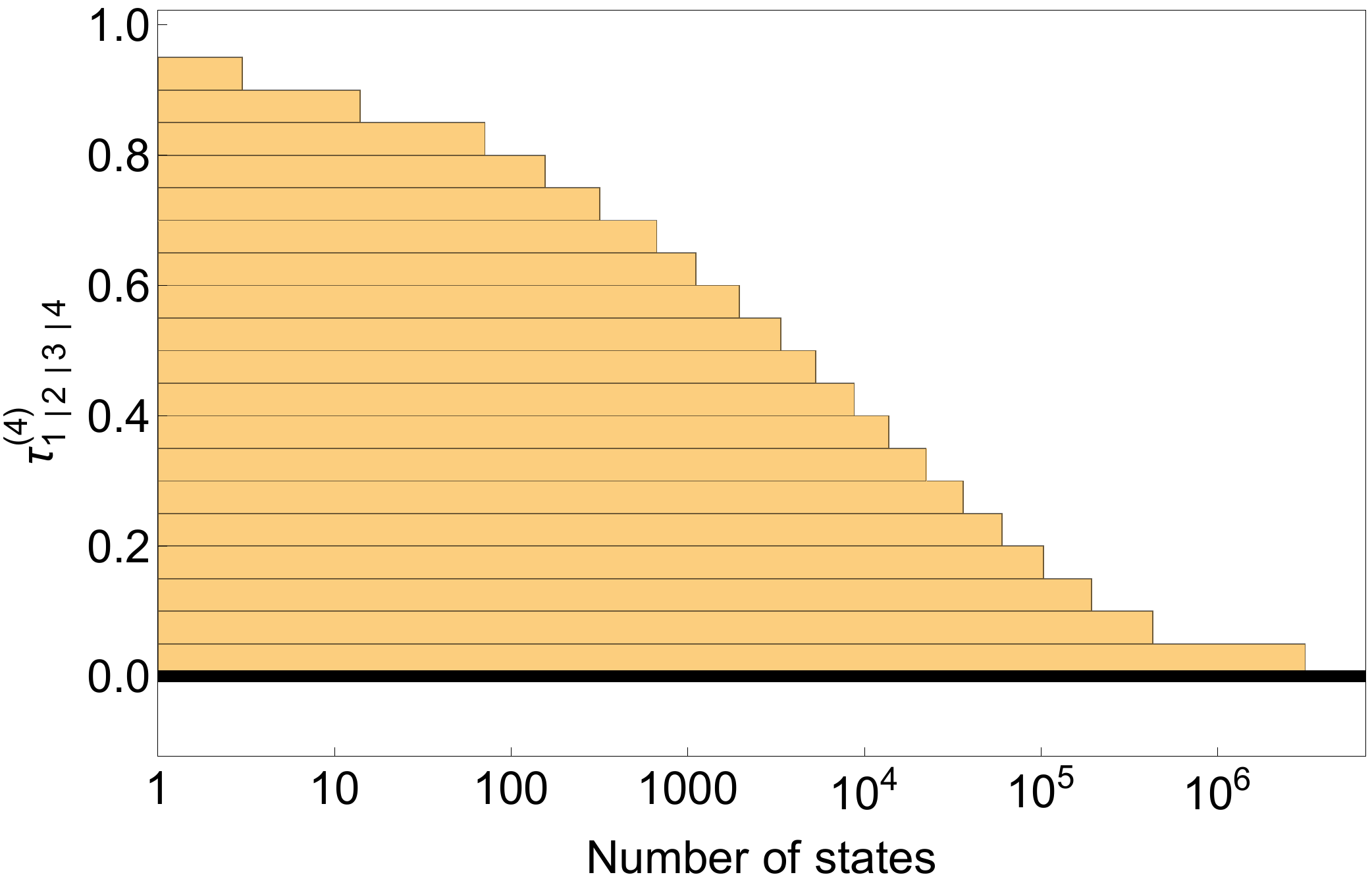}}
    \caption{A numerical analysis of $5\times 10^5$ states in each SLOCC class with a randomized choice of the complex parameters $a,b,c,d$ and the SLOCC transformations $A_k$. The figure shows a logarithmically scaled histogram of randomly generated pure four-qubit states with the corresponding lower bound of the residual $\tau_{1|2|3|4}^{(4)}$ obtained using the zero-$E$ approximation \cite{rodriques_2014}. The residual is defined by: (a) the natural strong monogamy inequality as in \eqref{eq:natural-sm} and \eqref{eq:natural-sm2}, (b) the modified strong monogamy inequality as in \eqref{eq:mu} with $\mu=3$, (c) the modified strong monogamy inequality as in \eqref{eq:q} with $q=4$.}
    \label{fig:hist}
\end{figure*}

\begin{figure*}
    \centering
    \subfloat[]{\includegraphics[width=5.9cm]{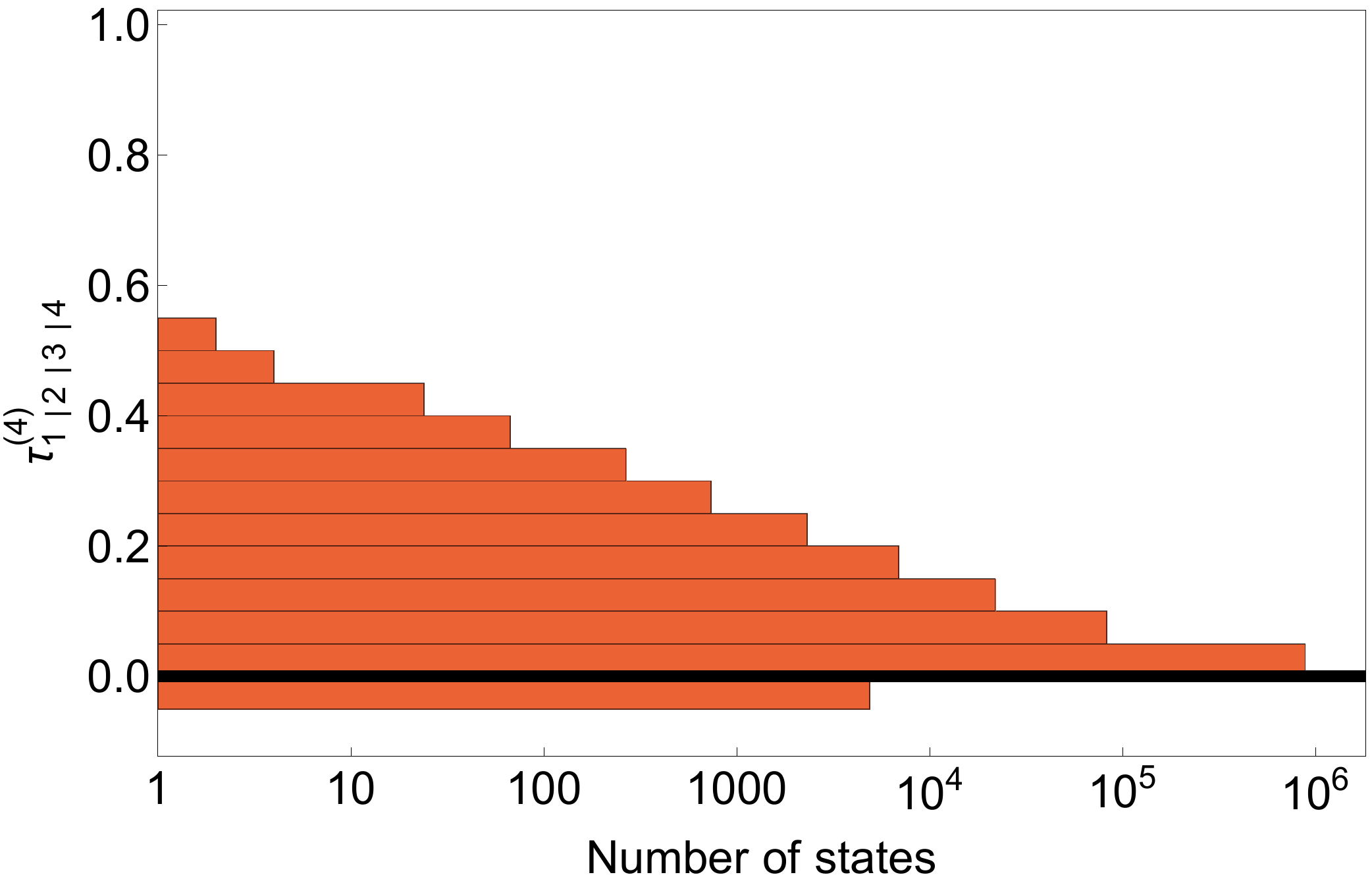}}
    \subfloat[]{\includegraphics[width=5.8cm]{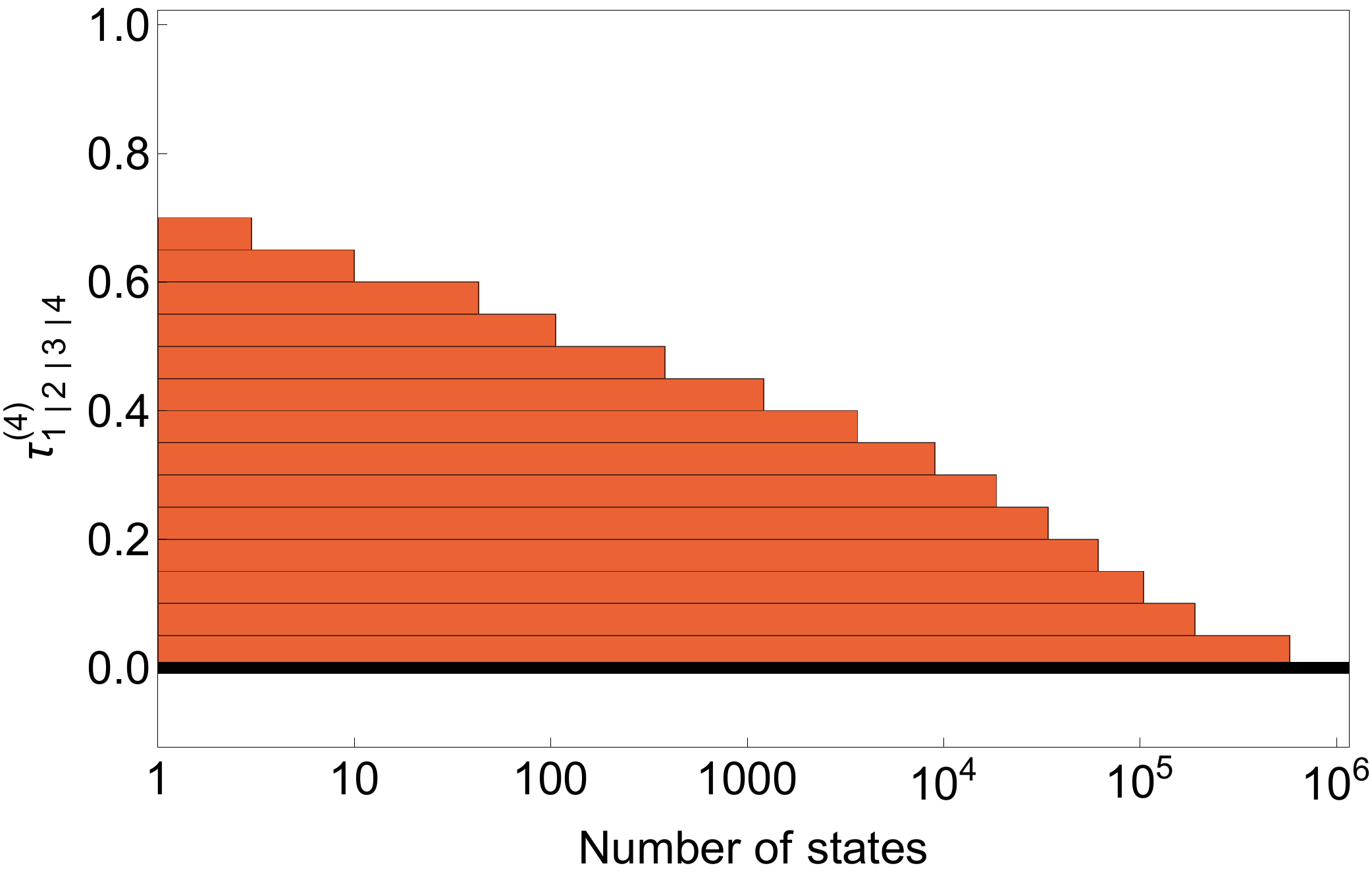}}
    \subfloat[]{\includegraphics[width=5.8cm]{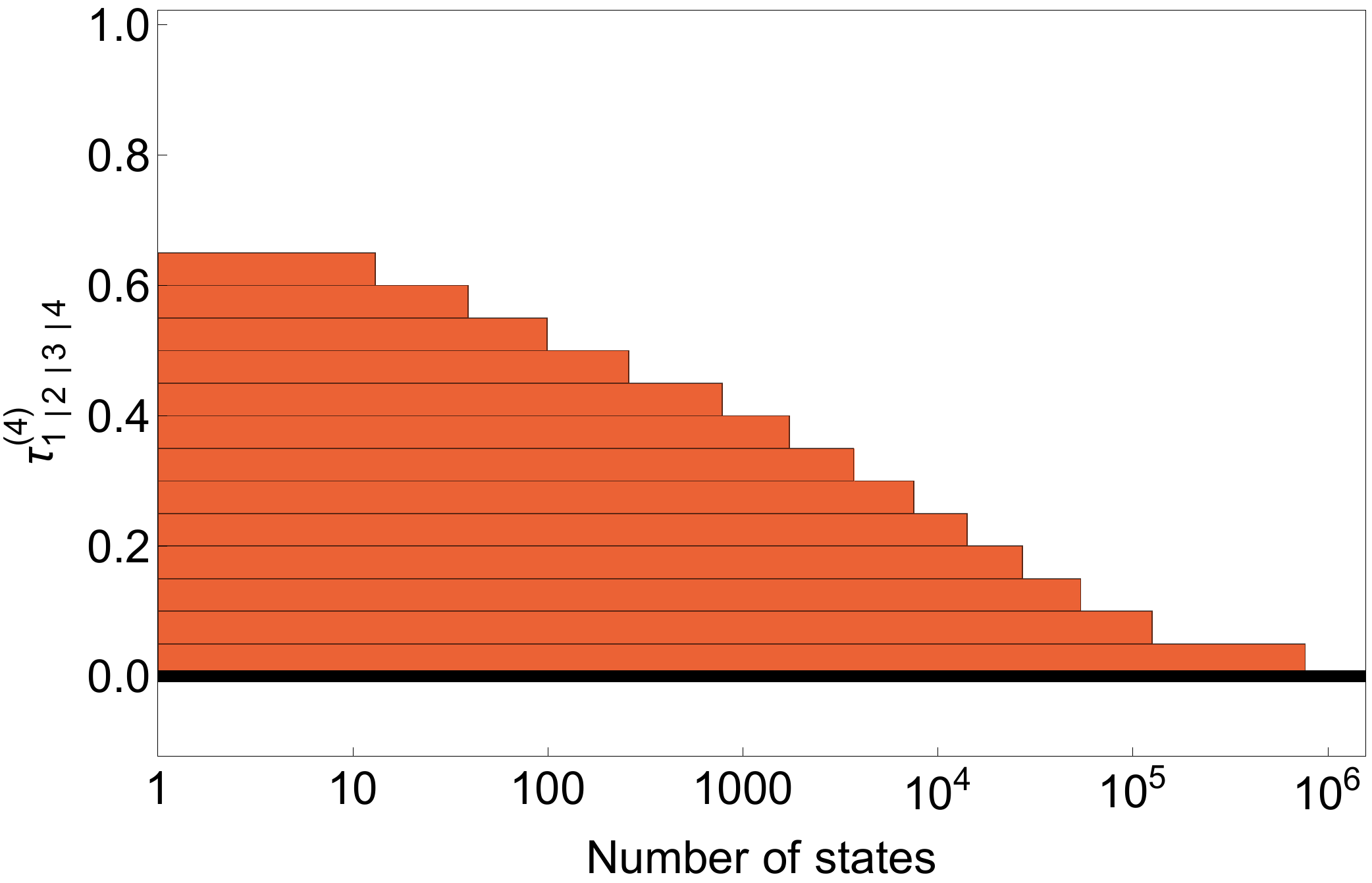}}
    \caption{A numerical analysis of $10^6$ states in the violating subclass of class 2, defined by the state $\ket{G^2_{a,b,c}}$ with $c=a$. The figure shows a logarithmically scaled histogram of randomly generated pure four-qubit states with the corresponding value of the residual $\tau_{1|2|3|4}^{(4)}$. The residual is defined by: (a) the natural strong monogamy inequality as in \eqref{eq:natural-sm} and \eqref{eq:natural-sm2}, (b) the modified strong monogamy inequality as in \eqref{eq:mu} with $\mu=3$, (c) the modified strong monogamy inequality as in \eqref{eq:q} with $q=4$. We note that the values in (a) and (b) are the exact values, while the residual in (c) uses the zero-$E$ approximation \cite{rodriques_2014} as before.}
    \label{fig:hist2}
\end{figure*}

The residual of the example state $\rho_{a_0,x}^2$ redefined this way is found to always be non-zero for $\mu \gtrsim 2.15$, and we present it in Fig.~\ref{fig:res}(b) for $\mu=3$. We recall that general results for the choice of $\mu=3$ in this approach have been reported in \cite{regula_2014}, where an extensive numerical test of $8\times 10^6$ states over all SLOCC classes provided evidence that this modified inequality should hold for all four-qubit states. We present detailed results of using this method on the residuals of randomized generic states in Fig.~\ref{fig:hist}(b), and specifically on states in the violating subclass---i.e.~the subclass of states which can exhibit violation of the strong monogamy inequality of Eq.~(\ref{eq:natural-sm})---in Fig.~\ref{fig:hist2}(b).

As the homogeneous degree of the measure in the convex roof remains unaffected, we retain the useful properties of the square root of the three-tangle, in particular the linear scaling of the convex roof \cite{viehmann_2012} as well as the one-root property \cite{regula_2016} where applicable. We note, however, that while this approach provides a straightforward method of rescaling the appropriate terms, it changes the overall degree of the inequality, which might be considered undesirable \cite{eltschka_2009}.

\subsection{Modifying the inequality: approach 2}

We can also consider a second alternative approach, in which the convex roof of the three-tangle is now taken to be
\begin{equation}
\label{eq:q}
    \tau_{i|j|k}^{(3)}(\rho) = \left[ \min_{\{p_n, \ket{\psi_n}\}} \sum_n p_n \left(\tau_{i|j|k}^{(3)}(\ket{\psi_n})\right)^{1/q} \right]^q\end{equation}
for some power $q \geq 2$. We note that $((\tau^{(3)}_{i|j|k}(\ket{\psi_n}))^{1/q}$ is an entanglement monotone for every $q\geq 1$ \cite{eltschka_2012} and now the degree of the inequality is left intact, but the degree of the measure in the convex roof is changed, which means the linear scaling of the convex roof no longer holds \cite{viehmann_2012} and the latter cannot be evaluated exactly by exploiting the one-root property \cite{regula_2016}. We therefore need to employ the zero-$E$ approximation \cite{rodriques_2014} to provide bounds on the residual for all SLOCC classes, including the degenerate subclasses of class 2. We find numerically that the lower bound for the residual of the example state $\rho_{a_0,x}^2$ is always non-zero for $q \gtrsim 2.28$, and we plot it in Fig.~\ref{fig:res}(c) for a choice of $q=4$. 

A numerical test of $1.5 \times 10^7$ randomly generated states over all SLOCC classes, and specifically an investigation of the degenerate subclasses of class 2, indicates that the modified monogamy inequality of this form might be satisfied in general when $q \gtrsim 2.42$. In particular, the inequality should clearly hold for the choice $q=4$, which would make the modified three-tangle a polynomial of degree 1 in the coefficients of the wave function. Because a choice of $2<q<4$ or $q>4$ would result in a non-integer scaling with the state coefficients, we consider this as the only possible value that $q$ can take. This can then be considered as the most straightforward and sensible generalization of the CKW monogamy inequality to four-qubit systems which we find likely to hold for all states.


\subsection{Discussion}

We now discuss and compare the two proposed modifications to the strong monogamy inequality in more detail.

The second approach can be seen to better reproduce the natural behavior of the straightforward extension of the CKW monogamy inequality of Eqs.~\eqref{eq:natural-sm} and \eqref{eq:natural-sm2}. Comparing Fig.~\ref{fig:res}(b) and \ref{fig:res}(c), we can see for the example state $\rho^2_{a_0,x}$ that approach 1 adds an artificial residual for all values of $x$, while approach 2 exhibits a similar behavior to Fig.~\ref{fig:res}(a) in the limiting cases, with $\tau_{1|2|3|4}^{(4)} \to 0$ as $x \to \pm \infty$.

To see how the two approaches affect the residual of states in general, we present the result of a numerical test of $4.5 \times 10^6$ states over all SLOCC classes in Fig.~\ref{fig:hist}. We see that both of the approaches retain the spread of the values $\tau_{1|2|3|4}^{(4)}$, with the biggest amount of states having very low values of the residual; however, an overall increase of the residual is more pronounced in approach 1. A similar comparison has been performed in Fig.~\ref{fig:hist2} for $10^6$ states in the degenerate subclass of SLOCC class 2, showing how the two approaches modify the residual of the states to eliminate all violations.

Additionally, let us remark that the class of generalized W states still saturates the strong monogamy inequality in both of the modified approaches, given that such states saturate the original CKW inequality \eqref{eq:osborne} as already noted in \cite{coffman_2000}, while all their $k$-partite entanglements (with $k>2$) are identically vanishing since the reduced density matrix of a generalized W state is a mixture of a product state and a generalized W state, whose mixed tangle is zero \cite{osterloh_2010} (see also Ref. \cite{kim_2014}).

We note that the residual tangle obtained here cannot be a SLOCC invariant \cite{eltschka_2009}. This contrasts with the three-qubit case where the residual three-tangle is a genuine three-partite entanglement measure invariant under SLOCC. We believe nonetheless that our proposed inequalities can still place important and intuitive constraints on the entanglement sharing among four (or more) parties.

The modified inequalities can be readily generalized to $n$ qubits, although the complex computational problems involved in calculating the convex roofs of entanglement measures for higher-rank states make it difficult to perform any kind of numerical analysis in larger systems. Even for five qubits, to lower-bound the 5-partite residual we need to obtain a lower bound on rank-4 mixed three-tangles, and no suitable bounds exist at the moment which provide sufficiently accurate results.

Finally, note that verification of a monogamy inequality for pure states implies, by convexity of the involved measures, its validity for general mixed states as well; this argument applies to all of the described approaches.

\section{Conclusions}

We have investigated the possibility of extending the standard CKW monogamy inequality \cite{coffman_2000} to four-qubit states, accounting also for the three-partite entanglement in the reduced three-qubit partitions. We have shown that the most natural and intuitive extension of the inequality, where the three-partite terms are counted on equal footing with the bipartite terms, does not hold in general. To this end, we have presented an analytical example of a family of states which violates this inequality, and for which the four-partite residual can be obtained exactly and be shown to be negative, confirming a conjecture raised in \cite{regula_2014}.

To resolve the violations of the extended monogamy inequality, we introduced two alternative methods to modify it. The methods were investigated both analytically on an example family of states as well as by an extensive numerical test of randomly generated states. It is not clear a priori which of the methods would represent a better choice to investigate the distribution of four-qubit entanglement, but our analysis suggests that tweaking the power of the three-tangle in the convex roof while retaining the overall degree of all terms of the the inequality (referred to as approach 2 in this paper) more closely resembles the properties of a desired strong monogamy extension. This suggests that an approach of this kind may indeed provide the most natural extension of the CKW inequality to more than three qubits, and that such an approach may hold in general, although it makes an analytical investigation much more difficult because it does not preserve some simplified characteristics which apply to the convex roof of the square root of the tangle \cite{regula_2016,viehmann_2012}.

Our analysis marks a step towards understanding the constraints of entanglement sharing in multipartite systems, but it also introduces several questions which are worth of consideration in future work. In particular, an analytical proof of a strong monogamy inequality holding in the most general case, already in the case of four qubits, would of course be necessary to complete our investigation. Without being able to investigate even the five-qubit case, we can only conjecture a generalized form of monogamy to hold, but our results provide strong evidence towards the existence of a strong monogamy constraint on arbitrary states of four qubits as manifested by a positive residual four-qubit term. It would be very interesting to answer the question of whether a simple class of states violating the natural extension of monogamy exists in systems of more qubits \cite{regula_2014} and higher-dimensional systems \cite{koashi_2004,adesso_2006,kim_2009} and to obtain a better understanding of their entanglement properties.

\acknowledgments{
We thank the European Research Council (ERC StG GQCOP, Grant No.~637352) for financial support. We acknowledge fruitful discussions with Sara {Di Martino}, Soojoon Lee, Barbara Kraus, Jens Siewert, and Frank Verstraete.}

\bibliography{main}

\begin{thebibliography}{39}%
\makeatletter
\providecommand \@ifxundefined [1]{%
 \@ifx{#1\undefined}
}%
\providecommand \@ifnum [1]{%
 \ifnum #1\expandafter \@firstoftwo
 \else \expandafter \@secondoftwo
 \fi
}%
\providecommand \@ifx [1]{%
 \ifx #1\expandafter \@firstoftwo
 \else \expandafter \@secondoftwo
 \fi
}%
\providecommand \natexlab [1]{#1}%
\providecommand \enquote  [1]{``#1''}%
\providecommand \bibnamefont  [1]{#1}%
\providecommand \bibfnamefont [1]{#1}%
\providecommand \citenamefont [1]{#1}%
\providecommand \href@noop [0]{\@secondoftwo}%
\providecommand \href [0]{\begingroup \@sanitize@url \@href}%
\providecommand \@href[1]{\@@startlink{#1}\@@href}%
\providecommand \@@href[1]{\endgroup#1\@@endlink}%
\providecommand \@sanitize@url [0]{\catcode `\\12\catcode `\$12\catcode
  `\&12\catcode `\#12\catcode `\^12\catcode `\_12\catcode `\%12\relax}%
\providecommand \@@startlink[1]{}%
\providecommand \@@endlink[0]{}%
\providecommand \url  [0]{\begingroup\@sanitize@url \@url }%
\providecommand \@url [1]{\endgroup\@href {#1}{\urlprefix }}%
\providecommand \urlprefix  [0]{URL }%
\providecommand \Eprint [0]{\href }%
\providecommand \doibase [0]{http://dx.doi.org/}%
\providecommand \selectlanguage [0]{\@gobble}%
\providecommand \bibinfo  [0]{\@secondoftwo}%
\providecommand \bibfield  [0]{\@secondoftwo}%
\providecommand \translation [1]{[#1]}%
\providecommand \BibitemOpen [0]{}%
\providecommand \bibitemStop [0]{}%
\providecommand \bibitemNoStop [0]{.\EOS\space}%
\providecommand \EOS [0]{\spacefactor3000\relax}%
\providecommand \BibitemShut  [1]{\csname bibitem#1\endcsname}%
\let\auto@bib@innerbib\@empty
\bibitem [{\citenamefont {Horodecki}\ \emph {et~al.}(2009)\citenamefont
  {Horodecki}, \citenamefont {Horodecki}, \citenamefont {Horodecki},\ and\
  \citenamefont {Horodecki}}]{Horodecki2009}%
  \BibitemOpen
  \bibfield  {author} {\bibinfo {author} {\bibfnamefont {R.}~\bibnamefont
  {Horodecki}}, \bibinfo {author} {\bibfnamefont {P.}~\bibnamefont
  {Horodecki}}, \bibinfo {author} {\bibfnamefont {M.}~\bibnamefont
  {Horodecki}}, \ and\ \bibinfo {author} {\bibfnamefont {K.}~\bibnamefont
  {Horodecki}},\ }\href {\doibase 10.1103/RevModPhys.81.865} {\bibfield
  {journal} {\bibinfo  {journal} {Rev Mod Phys}\ }\textbf {\bibinfo {volume}
  {81}},\ \bibinfo {pages} {865} (\bibinfo {year} {2009})}\BibitemShut
  {NoStop}%
\bibitem [{\citenamefont {Eltschka}\ and\ \citenamefont
  {Siewert}(2014)}]{eltschka_2014-1}%
  \BibitemOpen
  \bibfield  {author} {\bibinfo {author} {\bibfnamefont {C.}~\bibnamefont
  {Eltschka}}\ and\ \bibinfo {author} {\bibfnamefont {J.}~\bibnamefont
  {Siewert}},\ }\href {\doibase 10.1088/1751-8113/47/42/424005} {\bibfield
  {journal} {\bibinfo  {journal} {J. Phys. A: Math. Theor.}\ }\textbf {\bibinfo
  {volume} {47}},\ \bibinfo {pages} {424005} (\bibinfo {year}
  {2014})}\BibitemShut {NoStop}%
\bibitem [{\citenamefont {Nielsen}\ and\ \citenamefont
  {Chuang}(2011)}]{nielsen_2011}%
  \BibitemOpen
  \bibfield  {author} {\bibinfo {author} {\bibfnamefont {M.~A.}\ \bibnamefont
  {Nielsen}}\ and\ \bibinfo {author} {\bibfnamefont {I.~L.}\ \bibnamefont
  {Chuang}},\ }\href@noop {} {\emph {\bibinfo {title} {Quantum {{Computation}}
  and {{Quantum Information}}: 10th {{Anniversary Edition}}}}},\ \bibinfo
  {edition} {10th}\ ed.\ (\bibinfo  {publisher} {{Cambridge University
  Press}},\ \bibinfo {address} {New York, NY, USA},\ \bibinfo {year}
  {2011})\BibitemShut {NoStop}%
\bibitem [{\citenamefont {Amico}\ \emph {et~al.}(2008)\citenamefont {Amico},
  \citenamefont {Fazio}, \citenamefont {Osterloh},\ and\ \citenamefont
  {Vedral}}]{amico_2008}%
  \BibitemOpen
  \bibfield  {author} {\bibinfo {author} {\bibfnamefont {L.}~\bibnamefont
  {Amico}}, \bibinfo {author} {\bibfnamefont {R.}~\bibnamefont {Fazio}},
  \bibinfo {author} {\bibfnamefont {A.}~\bibnamefont {Osterloh}}, \ and\
  \bibinfo {author} {\bibfnamefont {V.}~\bibnamefont {Vedral}},\ }\href
  {\doibase 10.1103/RevModPhys.80.517} {\bibfield  {journal} {\bibinfo
  {journal} {Rev. Mod. Phys.}\ }\textbf {\bibinfo {volume} {80}},\ \bibinfo
  {pages} {517} (\bibinfo {year} {2008})}\BibitemShut {NoStop}%
\bibitem [{\citenamefont {Coffman}\ \emph {et~al.}(2000)\citenamefont
  {Coffman}, \citenamefont {Kundu},\ and\ \citenamefont
  {Wootters}}]{coffman_2000}%
  \BibitemOpen
  \bibfield  {author} {\bibinfo {author} {\bibfnamefont {V.}~\bibnamefont
  {Coffman}}, \bibinfo {author} {\bibfnamefont {J.}~\bibnamefont {Kundu}}, \
  and\ \bibinfo {author} {\bibfnamefont {W.~K.}\ \bibnamefont {Wootters}},\
  }\href {\doibase 10.1103/PhysRevA.61.052306} {\bibfield  {journal} {\bibinfo
  {journal} {Phys. Rev. A}\ }\textbf {\bibinfo {volume} {61}},\ \bibinfo
  {pages} {052306} (\bibinfo {year} {2000})}\BibitemShut {NoStop}%
\bibitem [{\citenamefont {Hill}\ and\ \citenamefont
  {Wootters}(1997)}]{hill_1997}%
  \BibitemOpen
  \bibfield  {author} {\bibinfo {author} {\bibfnamefont {S.}~\bibnamefont
  {Hill}}\ and\ \bibinfo {author} {\bibfnamefont {W.~K.}\ \bibnamefont
  {Wootters}},\ }\href {\doibase 10.1103/PhysRevLett.78.5022} {\bibfield
  {journal} {\bibinfo  {journal} {Phys. Rev. Lett.}\ }\textbf {\bibinfo
  {volume} {78}},\ \bibinfo {pages} {5022} (\bibinfo {year}
  {1997})}\BibitemShut {NoStop}%
\bibitem [{\citenamefont {Wootters}(1998)}]{wootters_1998}%
  \BibitemOpen
  \bibfield  {author} {\bibinfo {author} {\bibfnamefont {W.~K.}\ \bibnamefont
  {Wootters}},\ }\href {\doibase 10.1103/PhysRevLett.80.2245} {\bibfield
  {journal} {\bibinfo  {journal} {Phys. Rev. Lett.}\ }\textbf {\bibinfo
  {volume} {80}},\ \bibinfo {pages} {2245} (\bibinfo {year}
  {1998})}\BibitemShut {NoStop}%
\bibitem [{\citenamefont {D{\"u}r}\ \emph {et~al.}(2000)\citenamefont
  {D{\"u}r}, \citenamefont {Vidal},\ and\ \citenamefont {Cirac}}]{dur_2000}%
  \BibitemOpen
  \bibfield  {author} {\bibinfo {author} {\bibfnamefont {W.}~\bibnamefont
  {D{\"u}r}}, \bibinfo {author} {\bibfnamefont {G.}~\bibnamefont {Vidal}}, \
  and\ \bibinfo {author} {\bibfnamefont {J.~I.}\ \bibnamefont {Cirac}},\ }\href
  {\doibase 10.1103/PhysRevA.62.062314} {\bibfield  {journal} {\bibinfo
  {journal} {Phys. Rev. A}\ }\textbf {\bibinfo {volume} {62}},\ \bibinfo
  {pages} {062314} (\bibinfo {year} {2000})}\BibitemShut {NoStop}%
\bibitem [{\citenamefont {Verstraete}\ \emph {et~al.}(2003)\citenamefont
  {Verstraete}, \citenamefont {Dehaene},\ and\ \citenamefont {{De
  Moor}}}]{verstraete_2003}%
  \BibitemOpen
  \bibfield  {author} {\bibinfo {author} {\bibfnamefont {F.}~\bibnamefont
  {Verstraete}}, \bibinfo {author} {\bibfnamefont {J.}~\bibnamefont {Dehaene}},
  \ and\ \bibinfo {author} {\bibfnamefont {B.}~\bibnamefont {{De Moor}}},\
  }\href {\doibase 10.1103/PhysRevA.68.012103} {\bibfield  {journal} {\bibinfo
  {journal} {Phys. Rev. A}\ }\textbf {\bibinfo {volume} {68}},\ \bibinfo
  {pages} {012103} (\bibinfo {year} {2003})}\BibitemShut {NoStop}%
\bibitem [{\citenamefont {Osterloh}\ and\ \citenamefont
  {Siewert}(2005)}]{osterloh_2005}%
  \BibitemOpen
  \bibfield  {author} {\bibinfo {author} {\bibfnamefont {A.}~\bibnamefont
  {Osterloh}}\ and\ \bibinfo {author} {\bibfnamefont {J.}~\bibnamefont
  {Siewert}},\ }\href {\doibase 10.1103/PhysRevA.72.012337} {\bibfield
  {journal} {\bibinfo  {journal} {Phys. Rev. A}\ }\textbf {\bibinfo {volume}
  {72}},\ \bibinfo {pages} {012337} (\bibinfo {year} {2005})}\BibitemShut
  {NoStop}%
\bibitem [{\citenamefont {Streltsov}\ \emph {et~al.}(2012)\citenamefont
  {Streltsov}, \citenamefont {Adesso}, \citenamefont {Piani},\ and\
  \citenamefont {Bru{\ss}}}]{streltsov_2012}%
  \BibitemOpen
  \bibfield  {author} {\bibinfo {author} {\bibfnamefont {A.}~\bibnamefont
  {Streltsov}}, \bibinfo {author} {\bibfnamefont {G.}~\bibnamefont {Adesso}},
  \bibinfo {author} {\bibfnamefont {M.}~\bibnamefont {Piani}}, \ and\ \bibinfo
  {author} {\bibfnamefont {D.}~\bibnamefont {Bru{\ss}}},\ }\href {\doibase
  10.1103/PhysRevLett.109.050503} {\bibfield  {journal} {\bibinfo  {journal}
  {Phys. Rev. Lett.}\ }\textbf {\bibinfo {volume} {109}},\ \bibinfo {pages}
  {050503} (\bibinfo {year} {2012})}\BibitemShut {NoStop}%
\bibitem [{\citenamefont {Lancien}\ \emph {et~al.}(2016)\citenamefont
  {Lancien}, \citenamefont {{Di Martino}}, \citenamefont {Huber}, \citenamefont
  {Piani}, \citenamefont {Adesso},\ and\ \citenamefont
  {Winter}}]{lancien_2016}%
  \BibitemOpen
  \bibfield  {author} {\bibinfo {author} {\bibfnamefont {C.}~\bibnamefont
  {Lancien}}, \bibinfo {author} {\bibfnamefont {S.}~\bibnamefont {{Di
  Martino}}}, \bibinfo {author} {\bibfnamefont {M.}~\bibnamefont {Huber}},
  \bibinfo {author} {\bibfnamefont {M.}~\bibnamefont {Piani}}, \bibinfo
  {author} {\bibfnamefont {G.}~\bibnamefont {Adesso}}, \ and\ \bibinfo {author}
  {\bibfnamefont {A.}~\bibnamefont {Winter}},\ }\href@noop {} {} (\bibinfo
  {year} {2016}),\ \Eprint {http://arxiv.org/abs/1604.02189} {arXiv:1604.02189
  [quant-ph]} \BibitemShut {NoStop}%
\bibitem [{\citenamefont {Osborne}\ and\ \citenamefont
  {Verstraete}(2006)}]{osborne_2006}%
  \BibitemOpen
  \bibfield  {author} {\bibinfo {author} {\bibfnamefont {T.~J.}\ \bibnamefont
  {Osborne}}\ and\ \bibinfo {author} {\bibfnamefont {F.}~\bibnamefont
  {Verstraete}},\ }\href {\doibase 10.1103/PhysRevLett.96.220503} {\bibfield
  {journal} {\bibinfo  {journal} {Phys. Rev. Lett.}\ }\textbf {\bibinfo
  {volume} {96}},\ \bibinfo {pages} {220503} (\bibinfo {year}
  {2006})}\BibitemShut {NoStop}%
\bibitem [{\citenamefont {Eltschka}\ \emph {et~al.}(2009)\citenamefont
  {Eltschka}, \citenamefont {Osterloh},\ and\ \citenamefont
  {Siewert}}]{eltschka_2009}%
  \BibitemOpen
  \bibfield  {author} {\bibinfo {author} {\bibfnamefont {C.}~\bibnamefont
  {Eltschka}}, \bibinfo {author} {\bibfnamefont {A.}~\bibnamefont {Osterloh}},
  \ and\ \bibinfo {author} {\bibfnamefont {J.}~\bibnamefont {Siewert}},\ }\href
  {\doibase 10.1103/PhysRevA.80.032313} {\bibfield  {journal} {\bibinfo
  {journal} {Phys. Rev. A}\ }\textbf {\bibinfo {volume} {80}},\ \bibinfo
  {pages} {032313} (\bibinfo {year} {2009})}\BibitemShut {NoStop}%
\bibitem [{\citenamefont {Gour}\ and\ \citenamefont
  {Wallach}(2010)}]{gour_2010-1}%
  \BibitemOpen
  \bibfield  {author} {\bibinfo {author} {\bibfnamefont {G.}~\bibnamefont
  {Gour}}\ and\ \bibinfo {author} {\bibfnamefont {N.~R.}\ \bibnamefont
  {Wallach}},\ }\href {\doibase 10.1063/1.3511477} {\bibfield  {journal}
  {\bibinfo  {journal} {J. Math. Phys.}\ }\textbf {\bibinfo {volume} {51}},\
  \bibinfo {pages} {112201} (\bibinfo {year} {2010})}\BibitemShut {NoStop}%
\bibitem [{\citenamefont {Cornelio}(2013)}]{cornelio_2013}%
  \BibitemOpen
  \bibfield  {author} {\bibinfo {author} {\bibfnamefont {M.~F.}\ \bibnamefont
  {Cornelio}},\ }\href {\doibase 10.1103/PhysRevA.87.032330} {\bibfield
  {journal} {\bibinfo  {journal} {Phys. Rev. A}\ }\textbf {\bibinfo {volume}
  {87}},\ \bibinfo {pages} {032330} (\bibinfo {year} {2013})}\BibitemShut
  {NoStop}%
\bibitem [{\citenamefont {Regula}\ \emph {et~al.}(2014)\citenamefont {Regula},
  \citenamefont {{Di Martino}}, \citenamefont {Lee},\ and\ \citenamefont
  {Adesso}}]{regula_2014}%
  \BibitemOpen
  \bibfield  {author} {\bibinfo {author} {\bibfnamefont {B.}~\bibnamefont
  {Regula}}, \bibinfo {author} {\bibfnamefont {S.}~\bibnamefont {{Di
  Martino}}}, \bibinfo {author} {\bibfnamefont {S.}~\bibnamefont {Lee}}, \ and\
  \bibinfo {author} {\bibfnamefont {G.}~\bibnamefont {Adesso}},\ }\href
  {\doibase 10.1103/PhysRevLett.113.110501} {\bibfield  {journal} {\bibinfo
  {journal} {Phys. Rev. Lett.}\ }\textbf {\bibinfo {volume} {113}},\ \bibinfo
  {pages} {110501} (\bibinfo {year} {2014})}\BibitemShut {NoStop}%
\bibitem [{\citenamefont {Eltschka}\ and\ \citenamefont
  {Siewert}(2015)}]{eltschka_2015}%
  \BibitemOpen
  \bibfield  {author} {\bibinfo {author} {\bibfnamefont {C.}~\bibnamefont
  {Eltschka}}\ and\ \bibinfo {author} {\bibfnamefont {J.}~\bibnamefont
  {Siewert}},\ }\href {\doibase 10.1103/PhysRevLett.114.140402} {\bibfield
  {journal} {\bibinfo  {journal} {Phys. Rev. Lett.}\ }\textbf {\bibinfo
  {volume} {114}},\ \bibinfo {pages} {140402} (\bibinfo {year}
  {2015})}\BibitemShut {NoStop}%
\bibitem [{\citenamefont {Adesso}\ and\ \citenamefont
  {Illuminati}(2007)}]{adesso_2007}%
  \BibitemOpen
  \bibfield  {author} {\bibinfo {author} {\bibfnamefont {G.}~\bibnamefont
  {Adesso}}\ and\ \bibinfo {author} {\bibfnamefont {F.}~\bibnamefont
  {Illuminati}},\ }\href {\doibase 10.1103/PhysRevLett.99.150501} {\bibfield
  {journal} {\bibinfo  {journal} {Phys. Rev. Lett.}\ }\textbf {\bibinfo
  {volume} {99}},\ \bibinfo {pages} {150501} (\bibinfo {year}
  {2007})}\BibitemShut {NoStop}%
\bibitem [{\citenamefont {Uhlmann}(1998)}]{uhlmann_1998}%
  \BibitemOpen
  \bibfield  {author} {\bibinfo {author} {\bibfnamefont {A.}~\bibnamefont
  {Uhlmann}},\ }\href {\doibase 10.1023/A:1009664331611} {\bibfield  {journal}
  {\bibinfo  {journal} {Open Sys. \& Inf. Dyn.}\ }\textbf {\bibinfo {volume}
  {5}},\ \bibinfo {pages} {209} (\bibinfo {year} {1998})}\BibitemShut {NoStop}%
\bibitem [{\citenamefont {Lohmayer}\ \emph {et~al.}(2006)\citenamefont
  {Lohmayer}, \citenamefont {Osterloh}, \citenamefont {Siewert},\ and\
  \citenamefont {Uhlmann}}]{lohmayer_2006}%
  \BibitemOpen
  \bibfield  {author} {\bibinfo {author} {\bibfnamefont {R.}~\bibnamefont
  {Lohmayer}}, \bibinfo {author} {\bibfnamefont {A.}~\bibnamefont {Osterloh}},
  \bibinfo {author} {\bibfnamefont {J.}~\bibnamefont {Siewert}}, \ and\
  \bibinfo {author} {\bibfnamefont {A.}~\bibnamefont {Uhlmann}},\ }\href
  {\doibase 10.1103/PhysRevLett.97.260502} {\bibfield  {journal} {\bibinfo
  {journal} {Phys. Rev. Lett.}\ }\textbf {\bibinfo {volume} {97}},\ \bibinfo
  {pages} {260502} (\bibinfo {year} {2006})}\BibitemShut {NoStop}%
\bibitem [{\citenamefont {Eltschka}\ \emph {et~al.}(2008)\citenamefont
  {Eltschka}, \citenamefont {Osterloh}, \citenamefont {Siewert},\ and\
  \citenamefont {Uhlmann}}]{eltschka_2008}%
  \BibitemOpen
  \bibfield  {author} {\bibinfo {author} {\bibfnamefont {C.}~\bibnamefont
  {Eltschka}}, \bibinfo {author} {\bibfnamefont {A.}~\bibnamefont {Osterloh}},
  \bibinfo {author} {\bibfnamefont {J.}~\bibnamefont {Siewert}}, \ and\
  \bibinfo {author} {\bibfnamefont {A.}~\bibnamefont {Uhlmann}},\ }\href
  {\doibase 10.1088/1367-2630/10/4/043014} {\bibfield  {journal} {\bibinfo
  {journal} {New J. Phys.}\ }\textbf {\bibinfo {volume} {10}},\ \bibinfo
  {pages} {043014} (\bibinfo {year} {2008})}\BibitemShut {NoStop}%
\bibitem [{\citenamefont {Jung}\ \emph {et~al.}(2009)\citenamefont {Jung},
  \citenamefont {Hwang}, \citenamefont {Park},\ and\ \citenamefont
  {Son}}]{jung_2009}%
  \BibitemOpen
  \bibfield  {author} {\bibinfo {author} {\bibfnamefont {E.}~\bibnamefont
  {Jung}}, \bibinfo {author} {\bibfnamefont {M.-R.}\ \bibnamefont {Hwang}},
  \bibinfo {author} {\bibfnamefont {D.~K.}\ \bibnamefont {Park}}, \ and\
  \bibinfo {author} {\bibfnamefont {J.-W.}\ \bibnamefont {Son}},\ }\href
  {\doibase 10.1103/PhysRevA.79.024306} {\bibfield  {journal} {\bibinfo
  {journal} {Phys. Rev. A}\ }\textbf {\bibinfo {volume} {79}},\ \bibinfo
  {pages} {024306} (\bibinfo {year} {2009})}\BibitemShut {NoStop}%
\bibitem [{\citenamefont {Siewert}\ and\ \citenamefont
  {Eltschka}(2012)}]{siewert_2012}%
  \BibitemOpen
  \bibfield  {author} {\bibinfo {author} {\bibfnamefont {J.}~\bibnamefont
  {Siewert}}\ and\ \bibinfo {author} {\bibfnamefont {C.}~\bibnamefont
  {Eltschka}},\ }\href {\doibase 10.1103/PhysRevLett.108.230502} {\bibfield
  {journal} {\bibinfo  {journal} {Phys. Rev. Lett.}\ }\textbf {\bibinfo
  {volume} {108}},\ \bibinfo {pages} {230502} (\bibinfo {year}
  {2012})}\BibitemShut {NoStop}%
\bibitem [{\citenamefont {Viehmann}\ \emph {et~al.}(2012)\citenamefont
  {Viehmann}, \citenamefont {Eltschka},\ and\ \citenamefont
  {Siewert}}]{viehmann_2012}%
  \BibitemOpen
  \bibfield  {author} {\bibinfo {author} {\bibfnamefont {O.}~\bibnamefont
  {Viehmann}}, \bibinfo {author} {\bibfnamefont {C.}~\bibnamefont {Eltschka}},
  \ and\ \bibinfo {author} {\bibfnamefont {J.}~\bibnamefont {Siewert}},\ }\href
  {\doibase 10.1007/s00340-011-4864-x} {\bibfield  {journal} {\bibinfo
  {journal} {Appl. Phys. B}\ }\textbf {\bibinfo {volume} {106}},\ \bibinfo
  {pages} {533} (\bibinfo {year} {2012})}\BibitemShut {NoStop}%
\bibitem [{\citenamefont {Eltschka}\ and\ \citenamefont
  {Siewert}(2012)}]{eltschka_2012-1}%
  \BibitemOpen
  \bibfield  {author} {\bibinfo {author} {\bibfnamefont {C.}~\bibnamefont
  {Eltschka}}\ and\ \bibinfo {author} {\bibfnamefont {J.}~\bibnamefont
  {Siewert}},\ }\href@noop {} {\bibfield  {journal} {\bibinfo  {journal} {Sci.
  Rep.}\ }\textbf {\bibinfo {volume} {2}} (\bibinfo {year} {2012})}\BibitemShut
  {NoStop}%
\bibitem [{\citenamefont {Regula}\ and\ \citenamefont
  {Adesso}(2016)}]{regula_2016}%
  \BibitemOpen
  \bibfield  {author} {\bibinfo {author} {\bibfnamefont {B.}~\bibnamefont
  {Regula}}\ and\ \bibinfo {author} {\bibfnamefont {G.}~\bibnamefont
  {Adesso}},\ }\href {\doibase 10.1103/PhysRevLett.116.070504} {\bibfield
  {journal} {\bibinfo  {journal} {Phys. Rev. Lett.}\ }\textbf {\bibinfo
  {volume} {116}},\ \bibinfo {pages} {070504} (\bibinfo {year}
  {2016})}\BibitemShut {NoStop}%
\bibitem [{\citenamefont {Kim}(2014)}]{kim_2014}%
  \BibitemOpen
  \bibfield  {author} {\bibinfo {author} {\bibfnamefont {J.~S.}\ \bibnamefont
  {Kim}},\ }\href {\doibase 10.1103/PhysRevA.90.062306} {\bibfield  {journal}
  {\bibinfo  {journal} {Phys. Rev. A}\ }\textbf {\bibinfo {volume} {90}},\
  \bibinfo {pages} {062306} (\bibinfo {year} {2014})}\BibitemShut {NoStop}%
\bibitem [{\citenamefont {Choi}\ and\ \citenamefont {Kim}(2015)}]{choi_2015}%
  \BibitemOpen
  \bibfield  {author} {\bibinfo {author} {\bibfnamefont {J.~H.}\ \bibnamefont
  {Choi}}\ and\ \bibinfo {author} {\bibfnamefont {J.~S.}\ \bibnamefont {Kim}},\
  }\href {\doibase 10.1103/PhysRevA.92.042307} {\bibfield  {journal} {\bibinfo
  {journal} {Phys. Rev. A}\ }\textbf {\bibinfo {volume} {92}},\ \bibinfo
  {pages} {042307} (\bibinfo {year} {2015})}\BibitemShut {NoStop}%
\bibitem [{\citenamefont {Karmakar}\ \emph {et~al.}(2016)\citenamefont
  {Karmakar}, \citenamefont {Sen}, \citenamefont {Bhar},\ and\ \citenamefont
  {Sarkar}}]{karmakar_2016}%
  \BibitemOpen
  \bibfield  {author} {\bibinfo {author} {\bibfnamefont {S.}~\bibnamefont
  {Karmakar}}, \bibinfo {author} {\bibfnamefont {A.}~\bibnamefont {Sen}},
  \bibinfo {author} {\bibfnamefont {A.}~\bibnamefont {Bhar}}, \ and\ \bibinfo
  {author} {\bibfnamefont {D.}~\bibnamefont {Sarkar}},\ }\href {\doibase
  10.1103/PhysRevA.93.012327} {\bibfield  {journal} {\bibinfo  {journal} {Phys.
  Rev. A}\ }\textbf {\bibinfo {volume} {93}},\ \bibinfo {pages} {012327}
  (\bibinfo {year} {2016})}\BibitemShut {NoStop}%
\bibitem [{\citenamefont {Verstraete}\ \emph {et~al.}(2002)\citenamefont
  {Verstraete}, \citenamefont {Dehaene}, \citenamefont {{De Moor}},\ and\
  \citenamefont {Verschelde}}]{verstraete_2002}%
  \BibitemOpen
  \bibfield  {author} {\bibinfo {author} {\bibfnamefont {F.}~\bibnamefont
  {Verstraete}}, \bibinfo {author} {\bibfnamefont {J.}~\bibnamefont {Dehaene}},
  \bibinfo {author} {\bibfnamefont {B.}~\bibnamefont {{De Moor}}}, \ and\
  \bibinfo {author} {\bibfnamefont {H.}~\bibnamefont {Verschelde}},\ }\href
  {\doibase 10.1103/PhysRevA.65.052112} {\bibfield  {journal} {\bibinfo
  {journal} {Phys. Rev. A}\ }\textbf {\bibinfo {volume} {65}},\ \bibinfo
  {pages} {052112} (\bibinfo {year} {2002})}\BibitemShut {NoStop}%
\bibitem [{\citenamefont {Chterental}\ and\ \citenamefont
  {Djokovic}(2007)}]{chterental_2007}%
  \BibitemOpen
  \bibfield  {author} {\bibinfo {author} {\bibfnamefont {O.}~\bibnamefont
  {Chterental}}\ and\ \bibinfo {author} {\bibfnamefont {D.~Z.}\ \bibnamefont
  {Djokovic}},\ }in\ \href@noop {} {\emph {\bibinfo {booktitle} {Linear
  {{Algebra Research Advances}}}}},\ \bibinfo {editor} {edited by\ \bibinfo
  {editor} {\bibfnamefont {G.~D.}\ \bibnamefont {Ling}}}\ (\bibinfo
  {publisher} {{Nova Science Publishers}},\ \bibinfo {address} {New York},\
  \bibinfo {year} {2007})\ pp.\ \bibinfo {pages} {133--167}\BibitemShut
  {NoStop}%
\bibitem [{\citenamefont {Spee}\ \emph {et~al.}(2016)\citenamefont {Spee},
  \citenamefont {de~Vicente},\ and\ \citenamefont {Kraus}}]{spee_2015}%
  \BibitemOpen
  \bibfield  {author} {\bibinfo {author} {\bibfnamefont {C.}~\bibnamefont
  {Spee}}, \bibinfo {author} {\bibfnamefont {J.~I.}\ \bibnamefont
  {de~Vicente}}, \ and\ \bibinfo {author} {\bibfnamefont {B.}~\bibnamefont
  {Kraus}},\ }\href {\doibase http://dx.doi.org/10.1063/1.4946895} {\bibfield
  {journal} {\bibinfo  {journal} {J. Math. Phys.}\ }\textbf {\bibinfo {volume}
  {57}},\ \bibinfo {pages} {052201} (\bibinfo {year} {2016})}\BibitemShut
  {NoStop}%
\bibitem [{\citenamefont {Rodriques}\ \emph {et~al.}(2014)\citenamefont
  {Rodriques}, \citenamefont {Datta},\ and\ \citenamefont
  {Love}}]{rodriques_2014}%
  \BibitemOpen
  \bibfield  {author} {\bibinfo {author} {\bibfnamefont {S.}~\bibnamefont
  {Rodriques}}, \bibinfo {author} {\bibfnamefont {N.}~\bibnamefont {Datta}}, \
  and\ \bibinfo {author} {\bibfnamefont {P.}~\bibnamefont {Love}},\ }\href
  {\doibase 10.1103/PhysRevA.90.012340} {\bibfield  {journal} {\bibinfo
  {journal} {Phys. Rev. A}\ }\textbf {\bibinfo {volume} {90}},\ \bibinfo
  {pages} {012340} (\bibinfo {year} {2014})}\BibitemShut {NoStop}%
\bibitem [{\citenamefont {Eltschka}\ \emph {et~al.}(2012)\citenamefont
  {Eltschka}, \citenamefont {Bastin}, \citenamefont {Osterloh},\ and\
  \citenamefont {Siewert}}]{eltschka_2012}%
  \BibitemOpen
  \bibfield  {author} {\bibinfo {author} {\bibfnamefont {C.}~\bibnamefont
  {Eltschka}}, \bibinfo {author} {\bibfnamefont {T.}~\bibnamefont {Bastin}},
  \bibinfo {author} {\bibfnamefont {A.}~\bibnamefont {Osterloh}}, \ and\
  \bibinfo {author} {\bibfnamefont {J.}~\bibnamefont {Siewert}},\ }\href
  {\doibase 10.1103/PhysRevA.85.022301} {\bibfield  {journal} {\bibinfo
  {journal} {Phys. Rev. A}\ }\textbf {\bibinfo {volume} {85}},\ \bibinfo
  {pages} {022301} (\bibinfo {year} {2012})}\BibitemShut {NoStop}%
\bibitem [{\citenamefont {Osterloh}\ and\ \citenamefont
  {Siewert}(2010)}]{osterloh_2010}%
  \BibitemOpen
  \bibfield  {author} {\bibinfo {author} {\bibfnamefont {A.}~\bibnamefont
  {Osterloh}}\ and\ \bibinfo {author} {\bibfnamefont {J.}~\bibnamefont
  {Siewert}},\ }\href@noop {} {\bibfield  {journal} {\bibinfo  {journal} {New
  J. Phys.}\ }\textbf {\bibinfo {volume} {12}},\ \bibinfo {pages} {075025}
  (\bibinfo {year} {2010})}\BibitemShut {NoStop}%
\bibitem [{\citenamefont {Koashi}\ and\ \citenamefont
  {Winter}(2004)}]{koashi_2004}%
  \BibitemOpen
  \bibfield  {author} {\bibinfo {author} {\bibfnamefont {M.}~\bibnamefont
  {Koashi}}\ and\ \bibinfo {author} {\bibfnamefont {A.}~\bibnamefont
  {Winter}},\ }\href {\doibase 10.1103/PhysRevA.69.022309} {\bibfield
  {journal} {\bibinfo  {journal} {Phys. Rev. A}\ }\textbf {\bibinfo {volume}
  {69}},\ \bibinfo {pages} {022309} (\bibinfo {year} {2004})}\BibitemShut
  {NoStop}%
\bibitem [{\citenamefont {Adesso}\ and\ \citenamefont
  {Illuminati}(2006)}]{adesso_2006}%
  \BibitemOpen
  \bibfield  {author} {\bibinfo {author} {\bibfnamefont {G.}~\bibnamefont
  {Adesso}}\ and\ \bibinfo {author} {\bibfnamefont {F.}~\bibnamefont
  {Illuminati}},\ }\href {\doibase 10.1142/S0219749906001852} {\bibfield
  {journal} {\bibinfo  {journal} {Int. J. Quantum Inform.}\ }\textbf {\bibinfo
  {volume} {04}},\ \bibinfo {pages} {383} (\bibinfo {year} {2006})}\BibitemShut
  {NoStop}%
\bibitem [{\citenamefont {Kim}\ \emph {et~al.}(2009)\citenamefont {Kim},
  \citenamefont {Das},\ and\ \citenamefont {Sanders}}]{kim_2009}%
  \BibitemOpen
  \bibfield  {author} {\bibinfo {author} {\bibfnamefont {J.~S.}\ \bibnamefont
  {Kim}}, \bibinfo {author} {\bibfnamefont {A.}~\bibnamefont {Das}}, \ and\
  \bibinfo {author} {\bibfnamefont {B.~C.}\ \bibnamefont {Sanders}},\
  }\href@noop {} {\bibfield  {journal} {\bibinfo  {journal} {Phys. Rev. A}\
  }\textbf {\bibinfo {volume} {79}},\ \bibinfo {pages} {012329} (\bibinfo
  {year} {2009})}\BibitemShut {NoStop}%
\end{thebibliography}%

\end{document}